%% file: main.tex
\documentclass[conference]{IEEEtran}
\input{Preamble}
\begin{document}


\title{Hidden in Plain Sight: Where Developers Confess Self-Admitted Technical Debt}

\author{Murali Sridharan$^1$, Mikel Robredo$^1$, Leevi Rantala$^1$, Matteo Esposito$^1$, Valentina Lenarduzzi$^1$, Mika Mantyla$^2$\\
$^1$University of Oulu  --- $^2$University of Helsinki
\\
murali.sridharan@oulu.fi;
mikel.robredomanero@oulu.fi;
leevi.rantala@oulu.fi;\\
matteo.esposito@oulu.fi; 
valentina.lenarduzzi@oulu.fi
mika.mantyla@helsinki.fi
}

\maketitle

\begin{abstract}
\emph{Context.} Detecting Self-Admitted Technical Debt (SATD) is crucial for proactive software maintenance. Previous research has primarily targeted detecting and prioritizing SATD, with little focus on the source code afflicted with SATD. Our goal in this work is to connect the SATD comments with source code constructs that surround them. 

\noindent\emph{Method.} We leverage the extensive SATD dataset PENTACET, containing code comments from over 9000 Java Open Source Software (OSS) repositories. We quantitatively infer where SATD most commonly occurs and which code constructs/statements it most frequently affects. 

\noindent\emph{Results and Conclusions}. Our large-scale study links over 225,000 SATD comments to their surrounding code, showing that SATD mainly arises in inline code near definitions, conditionals, and exception handling, where developers face uncertainty and trade-offs, revealing it as an intentional signal of awareness during change rather than mere neglect.
\end{abstract}

\begin{IEEEkeywords}
Technical Debt, SATD, SATD patterns, Software Maintenance, Documentation
\end{IEEEkeywords}

\section{Introduction}
\input{sections/introduction}

\section{Background}
\label{sec:background}
\input{sections/background}

\section{Related Work}
\label{sec:related_work}

\input{sections/related_work}

\section{Empirical Study Design}
\label{sec:methodology}
\input{sections/methodology}


\section{Results }
\label{sec:results}
\input{sections/results}

\section{Discussion}
\label{sec:discussions}
\input{sections/discussions}

\section{Threats to Validity}
\label{sec:t2v}
\input{sections/validity}

\section{Conclusion}
\label{sec:conclusion}
\input{sections/conclusion}

\section*{Data Availability Statement}
We provide a replication package containing the raw data, the analysis scripts, and the results\footnote{\url{https://doi.org/10.5281/zenodo.10964573}}.

\balance
\bibliographystyle{IEEEtran}
\bibliography{main}
\end{document}

%% file: Preamble.tex
\IEEEoverridecommandlockouts
\usepackage{cite}
\usepackage{amsmath,amssymb,amsfonts}
\usepackage{algorithmic}
\usepackage{textcomp}
\usepackage{makecell}
\usepackage{tcolorbox}
\usepackage{listings}
\usepackage{xcolor}
\usepackage{multirow}
\usepackage{graphicx}
\usepackage{array}
\usepackage{scalefnt}
\usepackage{hyperref}
\usepackage{booktabs}
\usepackage{tabularray}
\UseTblrLibrary{booktabs}
\usepackage{todonotes}
\usepackage{xurl}
\usepackage{balance}
\usepackage{flushend}
\usepackage{threeparttable}
\def\BibTeX{{\rm B\kern-.05em{\sc i\kern-.025em b}\kern-.08em
    T\kern-.1667em\lower.7ex\hbox{E}\kern-.125emX}}

\definecolor{NavyBlue}{RGB}{0, 0, 128} 
\definecolor{CustomMagenta}{RGB}{255, 0, 255}
\definecolor{Maroon}{RGB}{128, 0, 0}
\definecolor{darkgreen}{rgb}{0,0.5,0}

\lstdefinestyle{papercode}{
  language=Java,
  basicstyle=\ttfamily\footnotesize,
  keywordstyle=\color{blue!70!black}\bfseries,
  stringstyle=\color{orange!60!black},
  commentstyle=\itshape\color{green!45!black},
  showstringspaces=false,
  tabsize=2,
  breaklines=true,
  breakatwhitespace=true,
  columns=fullflexible,
  keepspaces=true,
  frame=single,
  frameround=tttt,
  rulecolor=\color{black!15},
  backgroundcolor=\color{gray!3},
  captionpos=b,
  numbers=none,
  linewidth=0.98\linewidth,
  xleftmargin=4.5pt,
  framexleftmargin=0pt,
  framexrightmargin=0pt,
  aboveskip=1em,
  belowskip=1em
}

\usepackage[table,xcdraw]{xcolor}
\usepackage{multicol}
\usepackage{enumitem}

\usepackage{fontawesome}

\definecolor{main}{HTML}{CFCFCF}  
\definecolor{sub}{HTML}{CFCFCF}   

\tcbset{
    sharp corners,           
    colback = white,         
    before skip = 0.2cm,     
    after skip = 0.5cm       
}

\newtcolorbox{boxC}{
    colback = sub,  
    boxrule = 0pt   
}

\newcounter{keyTakeAwaysCounter} 
\newenvironment{keyTakeAways}[1][Key Take Away]{
    \addtocounter{keyTakeAwaysCounter}{1} 
    \begin{boxC}
    \faLightbulbO ~ \thekeyTakeAwaysCounter. \textbf{#1}.\\
}{
    \end{boxC}
}

\newcounter{keyRQAnswerCounter}
\newenvironment{keyRQAnswer}[1][RQAnswer]
    {
        \addtocounter{keyRQAnswerCounter}{1}
        \begin{boxC}{\faKey ~ \thekeyRQAnswerCounter. #1}
    }{
     \end{boxC}
    }
\def\BibTeX{{\rm B\kern-.05em{\sc i\kern-.025em b}\kern-.08em
    T\kern-.1667em\lower.7ex\hbox{E}\kern-.125emX}}

\renewcommand{\arraystretch}{1.4}

%% file: sections/introduction.tex
Technical Debt (TD), often resulting from expedient yet unsustainable development practices \cite{cunningham1992wycash}, has been extensively studied across multiple dimensions, including its identification, measurement, monitoring, prioritization, repayment, and documentation \cite{li2015systematic}. Self-Admitted Technical Debt (SATD) has gained particular attention since Potdar et al. \cite{potdar2014exploratory} revealed that developers often disclose intentional or temporary compromises directly in their source code comments.
Recent studies have advanced automated detection of SATD using methods ranging from keyword matching and text mining to sophisticated machine learning approaches \cite{huang2018identifying, de2020identifying, sridharan2021data}.


However, despite notable progress in detecting SATD, we still lack a clear understanding of the context in which developers introduce it. Source code and comments form two intertwined layers of documentation: comments convey a developer’s intent, rationale, or awareness of shortcuts, while the surrounding code embodies the practical manifestation of these decisions. Studying SATD from a textual perspective leaves unanswered questions about where in the code such admissions arise and why they tend to cluster in certain locations.
Understanding this contextual relationship is essential for both researchers and practitioners~\cite{alhefdhi2022framework, alomar2022satdbailiff, maldonado2017empirical, liu2018satd}. For researchers, it provides insights into the structural characteristics that accompany technical debt. This helps to refine predictive models of SATD and reduce false positives in automated detection. For practitioners, identifying the code constructs most associated with SATD can inform more effective debt monitoring, code review, and refactoring prioritization.
Recent studies suggest that specific syntactic constructs, such as conditionals, branching, and exception handling, tend to accumulate technical debt more frequently~\cite{alomar2022satdbailiff, maldonado2017empirical, liu2018satd}. This indicates that these regions may involve uncertainty or design trade-offs. However, the precise structural patterns linking SATD comments to their surrounding code remain largely unexplored. Without systematically analyzing these relationships, our understanding of how technical debt manifests and propagates in source code will remain incomplete.

\textbf{We aim} to bridge this gap by providing a systematic characterization of the source code that surrounds SATD occurrences. Specifically, our study investigates where SATD comments are most prevalent within projects, what types of code statements typically precede and follow them, and which recurring code patterns are most commonly associated with SATD instances. To the best of our knowledge, this is the first large-scale effort to operationalize and empirically analyze the source code context of SATD using a combination of structural and statistical perspectives. Our study provide the following \textbf{contributions}:

\begin{itemize}
\item Large-scale characterization of SATD-affected source code.
\item Framework and metrics for analyzing local SATD code context.
\item Identification of recurring SATD-related code patterns for research and practice.
\end{itemize}

Our large-scale analysis of over 225,000 SATD instances from 9,000 Java projects reveals that SATD is predominantly concentrated in non-header (inline) code, where developers make implementation-level decisions. SATD most frequently co-occurs with definitions, conditionals, branching, and exception handling, highlighting code regions of high uncertainty and trade-offs. Our findings show that SATD often arises just before new functionality or refactoring, suggesting it reflects developer awareness during change rather than neglect.

\noindent\textbf{Paper Structure.}  
In Section~\ref{sec:background}, we provide the theoretical background for the study. Section~\ref{sec:methodology} presents the study design, Section~\ref{sec:results} presents the results, and Section~\ref{sec:discussions} discusses them. Section~\ref{sec:related_work} discusses related work, and Section~\ref{sec:t2v} focuses on threats to the validity of our study, and Section~\ref{sec:conclusion}, draws the conclusions. 

%% file: sections/background.tex
\subsection{SATD Comments}
Source code comments, recognized as the second most used artifact for program comprehension after the code itself \cite{de2005study}, play a crucial role in understanding and maintaining software systems \cite{woodfield1981effect,tenny1988program,miyake2017replicated}. Over time, researchers have examined these comments across multiple dimensions: content, purpose, timing, and audience \cite{padioleau2009listening}.

Early classifications in Java identified four types of comments: object, style, type, and quality \cite{haouari2011good}. Later, broader taxonomies were proposed, distinguishing between copyright, header, member, inline, section, code, and task comments \cite{steidl2013quality}, and further refined to support maintainability \cite{pascarella2017classifying}. In this work, we adopt two high-level categories: Header and Non-Header comments. This distinction enables us to empirically characterize where SATD comments appear and to analyze their surrounding code context.
\textbf{Header comments} \cite{steidl2013quality} precede major code constructs such as classes or methods and describe their purpose or functionality \cite{sridhara2010towards,vassallo2014codes}. They typically serve documentation purposes, summarizing what follows.

\begin{lstlisting}[style=papercode, caption={Header comment for a simple Java class}, label={lst:book-header}]
/** Simple Book class implementation */
public class Book {
    private String title;
    private String author;

    // Member functions
    public String getTitle() {
        return title;
    }
}
\end{lstlisting}

An \textbf{header comment’s context} is limited to the construct it precedes. In the example above, the comment on line 1 describes the entire class, while the one on line 6 applies to the following method.

\textbf{Non-Header comments} \cite{malik2008understanding,chen2019automatically} appear within code bodies, often next to specific statements or logic blocks. They explain complex operations, clarify intent, or note improvements or fixes.

\begin{lstlisting}[style=papercode, caption={Example of Non-Header comments in Java code}, label={lst:nonheader-example}]
public ResponseEntity<String> uploadAirport() {
    // validate file
    if (file.isEmpty())
        // seek valid file
        return new ResponseEntity<>("Please upload a valid file.", BAD_REQUEST);
}

public String getAuthor() {
    // returns author name
    return author;
    // contains both first and last name
}
\end{lstlisting}

Unlike Header comments, \textbf{Non-Header comments code context} are influenced by both preceding and succeeding code. For instance, in the first method, the comment “// validate file” refers to the logic immediately following, while in the second, the final comment relates to the entire method’s return statement.

\subsection{Normalizing SATD Density by the Locality of SATD Comments}

Relying solely on raw counts of SATD comments can yield misleading conclusions, as comment placement varies substantially across different parts of the codebase. For instance, header comments may appear systematically in specific files, while inline or non-header comments are often scattered throughout the implementation. To account for such uneven distributions, we introduce normalization metrics that adjust for comment locality within the source code.
\textbf{Definition.} 
Let $\text{SATD}_{L}$ denote the number of SATD comments observed at a given code location $L$, and $\text{CMT}_{L}$ the total number of comments (SATD or not) at the same location. The \textit{Normalized SATD count by Location} metric, $N_{L}$, quantifies the relative proportion of SATD comments within that locality:

\begin{equation}
N_{L} = \frac{\text{SATD}_{L}}{\text{CMT}_{L}}
\end{equation}

Here, $L \in \{\text{Header}, \text{Non-Header}\}$, allowing us to separately evaluate SATD concentration in header and non-header regions. This formulation enables comparisons that are independent of the raw comment density.

\textbf{Practical Computation.}
To instantiate the metric, we compute two specific measures:

\begin{equation}
N_{\text{H}} = \frac{\text{SATD}_{\text{H}}}{\text{CMT}_{\text{H}}}
\qquad \text{and} \qquad
N_{\text{NH}} = \frac{\text{SATD}_{\text{NH}}}{\text{CMT}_{\text{NH}}}
\label{eq:instantiation}
\end{equation}

where $\text{H}$ and $\text{NH}$ denote header and non-header comments, respectively. Both ratios describe the share of SATD within their respective categories, ensuring that the analysis reflects comment behavior rather than raw frequency.

\textbf{Interpretation.}
Higher values of $N_{\text{H}}$ indicate a greater concentration of SATD within header comments, often associated with documentation or design intent, while elevated $N_{\text{NH}}$ values highlight technical debt occurring in more localized, implementation-level commentary. Together, these metrics provide a normalized lens through which SATD localization can be systematically compared across projects.

\subsection{Normalizing SATD Density by Code Context: Preceding and Succeeding Statements}

Beyond comment locality, it is essential to understand how the surrounding code context relates to the presence of SATD. Specifically, the types of statements that appear immediately before or after a comment can influence whether that comment expresses technical debt. To capture this phenomenon, we define normalization metrics that adjust SATD frequency by the structural characteristics of its adjacent statements.

\textbf{Definition.}
Let $\text{SATD}_{P}$ denote the number of SATD comments associated with a given \textit{preceding} statement type $P$, and $\text{CMT}_{P}$ represent the total number of comments (SATD or not) linked to that same statement type.  
The \textit{Normalized SATD by Preceding Statement Type} metric, $N_{P}$, is thus expressed as:

\begin{equation}
N_{P} = \frac{\text{SATD}_{P}}{\text{CMT}_{P}}
\label{eq:normalized_preceding}
\end{equation}

Similarly, let $\text{SATD}_{S}$ represent the number of SATD comments whose \textit{succeeding} statement type is $S$, and $\text{CMT}_{S}$ the total number of comments associated with that succeeding statement type.  
The \textit{Normalized SATD by Succeeding Statement Type} metric, $N_{S}$, is defined as:

\begin{equation}
N_{S} = \frac{\text{SATD}_{S}}{\text{CMT}_{S}}
\label{eq:normalized_succeeding}
\end{equation}

\textbf{Interpretation.}
These two measures enable comparisons across diverse code structures, such as definitions, conditionals, loops, or expressions, revealing where SATD is most concentrated relative to the general commenting behavior. By normalizing over all comment instances of a given statement type, the metrics isolate how much a particular syntactic context tends to co-occur with SATD, rather than simply reflecting the underlying code density.

\textbf{Practical Computation.}
In practice, both $N_{P}$ and $N_{S}$ are computed for each statement category defined in Section~\ref{sec:dc}, separately for header and non-header comments. This allows consistent cross-project comparisons and highlights which contextual statement types are more frequently associated with SATD occurrences, independently of project size or comment volume.

\subsection{Normalizing SATD Patterns Across Categories}

While the previous metrics normalize SATD occurrences relative to all comments or contextual statements, it is also important to analyze patterns within SATD instances themselves. This enables a focused examination of which types of code statements most frequently co-occur with SATD, revealing internal regularities in how developers express or surround technical debt.

\textbf{Definition.}
Let $\text{SATD}_{C}$ denote the number of SATD comments associated with all statement types belonging to a given category $C$, and let $\text{SATD}_{\text{ALL}}$ represent the total number of SATD comments across all categories defined in Section~\ref{sec:dc}.  
We define the \textit{Normalized SATD Pattern} metric, $N_{C}$, as:

\begin{equation}
N_{C} = \frac{\text{SATD}_{C}}{\text{SATD}_{\text{ALL}}}
\label{eq:normalized_pattern}
\end{equation}

Here, $C$ corresponds to any syntactic category (e.g., definitions, conditionals, expressions, or loops) used to group statement types. The metric, therefore, captures the relative share of SATD comments that fall within each structural category.

\textbf{Interpretation.}
Unlike the previous normalizations that compare SATD against all comments, $N_{C}$ focuses exclusively on the distribution of SATD itself. This allows for a direct assessment of which categories are most frequently affected by SATD and to what extent these patterns recur across projects. In essence, it normalizes within the domain of SATD comments only, highlighting their internal structure rather than their frequency relative to non-SATD commentary.

\textbf{Practical Computation.}
For each category listed in Section~\ref{sec:dc}, we compute $N_{C}$ to obtain the normalized frequency of SATD occurrences. These values are then used to identify the most recurrent SATD statement patterns, typically, the top three categories with the highest normalized proportions, providing an empirical view of how SATD clusters around specific syntactic contexts.

Finally, it is worth noticing that the normalization approach used in this study aligns with established software metrics methodology. In software analytics, raw metric counts are routinely normalized by a relevant size factor, such as lines of code or total comments, to enable fair, size-independent comparisons across systems or code regions. For example, defect density (defects per KLOC) and comment density (comment-to-code ratio) are long-standing normalized measures in software quality assessment \cite{fenton1998software, basili2002validation}. Following the same rationale, normalizing SATD counts by all comment or statement occurrences provides proportion-based indicators that reflect relative concentration rather than absolute frequency.

%% file: sections/related_work.tex
\input{Tables/RW}
SATD was first coined by~\cite{potdar2014exploratory}, who manually analyzed thousands of code comments and identified recurrent textual patterns (e.g., “TODO”, “FIXME”) that indicate intentional technical compromises. Following this seminal work, researchers extensively studied SATD detection as a text classification problem. Early approaches relied on simple keyword matching or text mining, while later studies applied machine learning and deep neural models to improve accuracy~\cite{potdar2014exploratory}. These techniques substantially advanced SATD classification, yet they largely treated code comments in isolation from the surrounding code. A consequence is high false-positive rates, many non-debt comments get misclassified as SATD because vital context is missing~\cite{potdar2014exploratory}. Recognizing this limitation, Yonekura et al.~\cite{Yonekura2025} integrated code context into a BERT-based model (CASTI) to detect SATD, significantly reducing false alarms and improving precision~\cite{Yonekura2025,potdar2014exploratory}. Such context-aware detection confirms that considering neighboring code can disambiguate a comment’s intent. However, these approaches focus on leveraging context for classifier performance, rather than characterizing what those contexts typically are. 

Our work addresses this gap by explicitly analyzing which code constructs tend to surround SATD comments, complementing prior detection studies with a systematic contextual perspective.

Beyond detection, researchers have also explored how SATD is managed and repaid in practice. Zampetti et al.~\cite{Zampetti2018} studied the removal of SATD comments across software evolution, finding that 20–50\% of SATD instances disappear “accidentally” when code containing them (e.g., whole classes or methods) is deleted. Notably, only a small fraction (~8\%) of SATD removals are explicitly mentioned in commit messages, suggesting most debt is resolved without fanfare. Their analysis further revealed that while some debts require extensive refactoring, many SATD fixes involve relatively localized changes, often to method calls or conditional logic. This indicates a link between SATD and specific code structures (e.g., conditional statements) even during remediation. Alhefdhi et al.~\cite{Alhefdhi2024} extended this line of inquiry towards automating SATD repayment. They developed a taxonomy of how SATD comments relate to the eventual code fixes (identifying five distinct categories of comment–code relationships) and designed a deep learning tool, DLRepay, to automatically generate “debt-free” code given a debt comment and its surrounding code. Their results demonstrate that SATD comments can indeed guide automated fixes in many cases, especially when the comment clearly describes a needed change. Compared to our work, which maps SATD to code context for understanding, Alhefdhi et al.~\cite{Alhefdhi2024}’s work uses the code context for transformation. Both underscore the importance of contextual links: one for repairing debt, and ours for characterizing how debt manifests in code.

In parallel, the community has produced tools and datasets to facilitate SATD research. SATDBailiff~\cite{alomar2022satdbailiff} is a notable tool that automates the mining and tracking of SATD comments over time. It uses a state-of-the-art classifier to detect SATD in code repositories, then monitors each detected comment through version history, recording any text updates and ultimately its removal. By reporting the lifespan and evolution of each debt item, SATDBailiff helps researchers and practitioners study how long debts survive and whether they get cleaned up or simply vanish via code deletion. However, while SATDBailiff excels at temporal tracking, it does not describe the code structures around the comments it flags. Our study instead provides a complementary spatial analysis of SATD, examining the immediate code context of debt comments rather than their history. On the data side, the field has progressed from small project-specific datasets to large-scale, context-rich corpora. The original SATD datasets (e.g., from Potdar’s and subsequent studies) contained tens of thousands of comments from a handful of Java projects. Recently, PENTACET~\cite{Sridharan2023} significantly expanded the scope by mining 9,000+ open-source Java projects and extracting 23 million code comments along with their surrounding code snippets. Over 250,000 of these comments are labeled as SATD, including not only the “easy-to-find” debt markers (like TODOs) but also more subtle debt comments identified via machine learning. PENTACET’s inclusion of preceding and succeeding code lines for each comment provides exactly the kind of contextual data that earlier SATD detectors lacked. In fact, our current study leverages this dataset to quantitatively link SATD comments with code constructs, a task that would have been impractical with prior smaller datasets. Furthermore, the emergence of cross-language SATD datasets is addressing the bias toward Java. Pham et al.~\cite{Pham2025} introduced CppSATD, the first extensive SATD dataset for C++ code, containing over 531,000 annotated comments with their code contexts. This dataset was motivated by the observation that most SATD research, including PENTACET, focused on Java, limiting generalizability across languages. By providing a rich resource for C++ with similar contextual information, CppSATD enables researchers to develop C++-specific debt detectors and to investigate whether SATD patterns in C++ mirror those in Java. Such resources (PENTACET for Java, CppSATD for C++) underscore a growing emphasis on context and breadth in SATD datasets, which in turn empowers studies like ours.

All in all, prior works (Table~\ref{tab:tools-datasets}) laid the foundation in identifying and removing SATD, and recent advances emphasize context-aware analysis, whether for improving detection accuracy, tracking debt over time, or enabling automated fixes. Our contribution builds on this momentum by systematically bridging the long-standing gap between SATD comments and the code they refer to. In contrast to pure classification efforts that treat comments as standalone text, we provide an explicit mapping of SATD to syntactic code elements (e.g., conditionals, declarations, exceptions). This novel angle offers complementary insights: where others have aimed to detect or resolve technical debt, we aim to describe and understand its local code context. By doing so, we enrich the knowledge of SATD’s nature and locality, which can inform more context-sensitive detection tools and guide developers in anticipating what kinds of code are most prone to incurring SATD.

%% file: Tables/RW.tex
\begin{table*}[t]
\centering
\caption{Related Works}
\label{tab:tools-datasets}
\resizebox{0.99\linewidth}{!}{%
\begin{tabular}{p{2cm}p{2cm}p{2cm}p{6.5cm}p{6.5cm}}
\toprule
\textbf{Reference} & \textbf{Type} & \textbf{Language / Scope} & \textbf{Main Contribution} & \textbf{Relation to Our Work} \\
\midrule
\textbf{SATDBailiff}~\cite{AlOmar2022} & Tool & Multi-language (Java, C\#) & Automates mining and longitudinal tracking of SATD across revisions; focuses on temporal debt lifespan and removal. & Focuses on \textit{temporal} tracking of SATD, whereas our work provides a \textit{spatial} characterization of its surrounding code constructs. \\
\textbf{PENTACET}~\cite{Sridharan2023} & Dataset & Java (9,000+ OSS repos) & 23M contextual comments; 250K SATD instances annotated with preceding and succeeding code. & Forms the empirical foundation of our study; we analyze SATD–code construct relationships at scale. \\
\textbf{CppSATD}~\cite{Pham2025} & Dataset & C++ (531K comments) & First large-scale C++ SATD dataset linking code context and comment semantics. & Extends SATD contextual research beyond Java; our methodology is aligned with this contextual linking direction. \\
\textbf{Zampetti et al. (2018)}~\cite{Zampetti2018} & Empirical Study & Java OSS & Explores whether SATD removals represent genuine debt repayment; highlights localized vs large-scale removal patterns. & Provides evidence that SATD often relates to specific code structures; our study systematizes this link. \\
\textbf{Alhefdhi et al. (2024)}~\cite{Alhefdhi2024} & Automated Repair Study & Multi-language & Introduces \textit{DLRepay}, a deep learning model for SATD repayment via comment–code relationship modeling. & Shares the focus on leveraging code context; we focus on characterizing rather than repairing context. \\
\textbf{Yonekura et al. (2025)}~\cite{Yonekura2025} & Detection Model & Java & CASTI: BERT-based SATD detector incorporating code context to reduce false positives. & Context is used for classification accuracy; our work uses it for interpretive characterization. \\
\bottomrule
\end{tabular}%
}
\end{table*}

%% file: sections/methodology.tex
In this section, we describe the study's design, following the guidelines by Wohlin et al. \cite{wohlin_experimentation_2024}. 

\subsection{Goal and Research Questions}
We formalized the \textbf{goal} of this study as follows: we aim to \textit{analyze and characterize} the source code context of SATD \textit{in order to understand} where SATD occurs, which code structures it most frequently coexists with, and whether recurring contextual patterns can be identified, from the \textit{viewpoint of} software quality researchers and practitioners, \textit{in the context of} large-scale open-source Java systems from the \texttt{PENTACET} dataset.
\begin{boxC}
\textbf{RQ$_1$}. Is SATD more prevalent in non-header comments than in header comments?
\end{boxC}
In large-scale codebases, developers often annotate SATD in diverse locations, ranging from file or class headers to inline implementation regions. The distinction between header and non-header comments reflects different abstraction levels of concern: header comments typically capture architectural or design debt, whereas non-header comments are more closely tied to implementation or maintenance debt~\cite{de2016investigating,de2016identifying,maldonado2017empirical}. Therefore, we are keen to investigate whether SATD predominantly occurs in non-header or header regions by measuring $N_{L}$ (Equation~\ref{eq:instantiation}) across projects and test the following \textbf{null} and \textbf{alternative} hypotheses:

\begin{itemize}
\setlength{\leftskip}{1em}
    \item[\textbf{H}$_{01}$:] \textit{There is no significant difference in SATD prevalence between header and non-header comments.}
    \item[\textbf{H}$_{11}$:] \textit{There is a significant difference in SATD prevalence between header and non-header comments.}
\end{itemize}

Understanding where SATD manifests most frequently provides the foundation to explore what kinds of code contexts surround SATD in each location. Hence, we ask:

\begin{boxC}
\textbf{RQ$_2$}. Do specific code statement types co-occur with SATD more often than others?
\end{boxC}
The code statements immediately surrounding SATD comments may reveal the technical context prompting self-admission. For instance, SATD near branching or exception statements might indicate decision or fault-handling motivation, whereas SATD around definitions or declarations could reflect design or initialization-related concerns. To quantitatively characterize the \textbf{code context} where SATD arises, we distinguish between high-level constructs (e.g., class or method definitions) and localized structures (e.g., loops, conditions, or expressions)~\cite{de2016identifying,maldonado2017empirical} and we test the following hypotheses:
\begin{itemize}
\setlength{\leftskip}{1em}
    \item[\textbf{H}$_{02}$:] \textit{No significant difference exists in the contextual frequency of SATD across code statement categories.}
    \item[\textbf{H}$_{12}$:] \textit{Specific categories (e.g., branching, exception handling) show higher SATD density, indicating contextual hotspots.}
\end{itemize}

Finally, understanding whether such a code statement may lead to recurring ''patterns" is paramount. Hence, we ask:

\begin{boxC}
\textbf{RQ$_3$}. Do recurring SATD patterns exist in non-header code?
\end{boxC}
Beyond individual contextual relationships, developers may repeatedly introduce SATD around characteristic combinations of preceding and succeeding code statements. Such recurring patterns can reveal structural tendencies in how technical debt manifests, offering valuable insights for predictive debt detection and IDE-based recommendations.

To uncover these recurring SATD-prone combinations, we analyze intra-debt frequencies across different statement pairings, focusing exclusively on non-header comments to isolate fine-grained contextual dependencies. To such aim, we conjecture the following hypotheses:
\begin{itemize}
\setlength{\leftskip}{1em}
    \item[\textbf{H}$_{03}$:] No specific pairing pattern occurs significantly more frequently among SATD instances.
    \item[\textbf{H}$_{13}$:] Specific code-comment combinations dominate SATD patterns.
\end{itemize}

\subsection{Context}
Our study builds on a large-scale dataset of open-source Java systems that captures how developers document and localize SATD in practice (\texttt{PENTACET}). The projects span diverse domains, sizes, and maturity levels, offering a broad view of real-world software development. Each SATD instance is linked to its precise location in the source code, enabling an analysis that connects natural language comments with their structural context. 

\subsection{Data Collection}
\label{sec:dc}
We base our analysis on the \texttt{PENTACET} dataset \cite{sridharan2023pentacet}, which includes more than 250,000 machine-annotated SATD comments collected from over 9,000 open-source Java repositories. Each comment in the dataset is linked to its surrounding code structure, extracted using the \textit{srcML} tool \cite{collard2013srcml}. The dataset provides a rich view of how developers document and localize technical debt across real-world projects.

Our focus lies on the code context around each SATD comment. We examine the statements that appear immediately before and after the comment. Across all projects, the dataset includes 58 unique statement types. To make the analysis systematic and readable, we group these statements into nine broader categories according to previous study suggestions~\cite{alhefdhi2022framework,sridharan2023pentacet}: Definitions (DEFN), Declarations (DECL), Expressions (EXPR), Conditionals (CNDTNL), Branching (BRNCH), Exceptions (EXCPTN), Loops (LOOPS), Documentation (DOC), and Miscellaneous (MISC).

We include statements that represent clear semantic or structural roles in the source code. We exclude ambiguous or generic constructs such as \textit{block}, \textit{parameter}, or \textit{EOF}, which do not convey meaningful context. We also classify \textit{import} and \textit{package} under the Definition category because they often occur at the file header and represent a definitional unit at the file level.

We focus only on the immediate context of each SATD comment that directly precedes and succeeds statements. We do not analyze nested or deeply embedded structures, since our goal is to capture the most explicit relationships between SATD comments and their surrounding code. 

\input{Tables/studycontext}

\subsection{Data Analysis}
This section presents the data analysis procedure to gather insights from the collected data.
We must assess the numerical metrics' distribution before choosing the proper correlation test \cite{falessi_enhancing_2023}. We tested \emph{\textbf{H$_{0\mathcal{N}}$} (data is normally distributed)} using Anderson-Darling (AD) test \cite{anderson1952asymptotic}. AD tests whether data points are sampled from a specific probability distribution, in this case, a normal distribution. We rejected the null hypothesis (H$_{0\mathcal{N}}$) for all the data; hence, we employed the \textit{Wilcoxon–Mann–Whitney U test} \cite{mann1947test} to assess whether the observed differences across categories were statistically significant. This non-parametric test is appropriate when comparing two independent groups with non-normal distributions, as it evaluates whether one group tends to produce higher values than the other. To estimate the magnitude of differences, we computed the \textit{rank-biserial correlation coefficient ($R$)} \cite{cureton1956rank}, interpreted as small ($0.1–0.24$), moderate ($0.24–0.37$), or large ($>$0.37) \cite{kampenes2007systematic}.

To answer \textbf{RQ1}, we investigated where SATD comments occur most frequently within the source code. We calculated the $N_{L}$ (Equation~\ref{eq:instantiation}) metric to measure the proportion of SATD in header and non-header regions across projects. We then compared these proportions using the Wilcoxon–Mann–Whitney U test to determine whether the difference in SATD prevalence between the two locations was statistically significant. The corresponding effect size ($R$) quantified the strength of the observed difference. We repeated this analysis across three project size categories—1–100 KLOC, $100–1000$ KLOC, and $>1000$ KLOC—to examine whether system scale influences localization trends.

To answer \textbf{RQ2}, we analyzed the code context surrounding SATD comments to identify which types of source code statements most often co-occur with them. We computed two complementary metrics: $N_P$ for the preceding statement and $N_S$ for the succeeding statement. Each metric normalizes the number of SATD-afflicted comments by the total number of comments associated with a specific statement type, enabling fair comparison across categories such as definitions, loops, conditionals, and exceptions. We again applied the Wilcoxon–Mann–Whitney U test and calculated the rank-biserial coefficient to identify which statement categories exhibit significantly higher SATD frequencies.

Finally, to answer \textbf{RQ3}, we explored whether recurring code-comment combinations emerge among non-header SATD instances. We used the $N_{C}$  metric, which measures the relative frequency of each pair of preceding and succeeding statement categories within the SATD subset. This normalization captures intra-SATD proportions and isolates the most frequent contextual pairings independent of general commenting practices. We then ranked these pairings to identify the top recurring SATD patterns that characterize how developers typically express technical debt in non-header code.

%% file: Tables/studycontext.tex
\begin{boxC}
\textbf{Statement Types In-Scope}
\\~\\
    \begin{minipage}[t]{0.45\linewidth}
\begin{description}[leftmargin=0.8em,labelsep=0.5em,style=sameline,font=\bfseries, nosep]
  \item[DEFN:] function, enum, class, interface, constructor, import, package
  \item[LOOPS:] do, for, while, range, while
  \item[EXPR:] expr\_stmt, call
  \item[BRNCH:] break, return, label, continue
 
\end{description}
\end{minipage}\hfill
\begin{minipage}[t]{0.45\linewidth}
\begin{description}[leftmargin=1em,labelsep=0.5em,style=sameline,font=\bfseries, nosep]
 \item[CNDTNL:] if, if\_stmt, else, case, switch, then, ternary, default, assert
  \item[EXCPTN:] catch, throw, try, finally
  \item[DECL:] decl\_stmt, function\_decl, constructor\_decl
  \item[DOC:] comment
\end{description}
\end{minipage}
\end{boxC}

\begin{boxC}
\textbf{Statement Types Excluded (Miscellaneous)}

\begin{minipage}[t]{\linewidth}
\begin{description}[leftmargin=1.2em,labelsep=0.5em,style=sameline,font=\bfseries, nosep]
  \item[MISC:] argument, argument\_list, block, block\_content, condition, index, name, operator, 
  parameter, parameter\_list, specifier, super, expr, decl, empty\_stmt, EOF, literal, annotation, annotation\_defn
\end{description}
\end{minipage}
\end{boxC}

%% file: sections/results.tex
In this Section, we report the results to answer our RQ. 

\subsection{SATD Localization ($RQ_1$)}
On average, 7.01\% of SATD comments are non-header as against 2.5\% header comments for projects of size 1-100 KLOC. For projects of size 100-1000 KLOC, 3.9\% of SATD comments are non-header, as against 0.9\% header comments, and for projects greater than 1000 KLOC, 1.9\% of non-header comments are SATD compared to 0.5\% header comments.  Consistently across all project sizes, non-header comments have approximately 3 to 4 times more SATD comments than header comments, demonstrating a consistent trend across varying project sizes (Table~\ref{tab:localization_statistical_test}).

\input{Tables/RQ1/characterization}

Across all project sizes, we could reject the null-hypothesis (p-value$<0.001$)  $H_{01}$ and affirm that \textit{there is a statistically significant difference in the prevalence of SATD comments between header and non-header} (Table~\ref{rq_1_mwstatistic}). Furthermore, the effect size ($R$ coefficient) indicates a large effect~\cite{kampenes2007systematic}, suggesting that non-header comments are consistently ranked higher in terms of SATD prevalence than header comments across project sizes. In conjunction with the normalized mean and median proportions, it is established that SATD comments are significantly prevalent in non-header comments. 
\input{Tables/RQ1/h01}
\begin{keyRQAnswer}[\textbf{SATD Distribution.}]
    SATD comments are significantly more prevalent and concentrated in non-header comments, independent of project size.
\end{keyRQAnswer}

\subsection{SATD Code Contextualization ($RQ_2$)}
Regarding Header comments, across all project sizes, `CLASS' comments have the highest SATD concentration with 23.22\% for 1-100 KLOC, 26.54\% for 100-1000 KLOC, and 30.15\% for over 1000 KLOC. Conversely, `FILE' comments have the lowest with 0.72\% for 1-100 KLOC, 4.53\% for 100-1000 KLOC, and 4.52\% for over 1000 KLOC (Table \ref{tab:h_succ_context} and Table \ref{tab:h_succ_context_2}). Hypothesis testing shows that differences in SATD between `CLASS' and other comments from `CONSTRUCTOR', `INTERFACE', and `FUNCTION'for each project size, respectively,  are significant with p-values less than 0.001, thus we can reject the null hypothesis in each case (Table~\ref{rq_2_1_mwstatistic}. Furthermore, we observe that `CLASS' header comments consistently contain a higher proportion of SATD across all project sizes. For Header comments, the effect size suggests that there is a large difference, indicating the frequency of `CLASS' as the succeeding code context of header comments is significantly high compared to other constructs, including `CONSTRUCTOR', `INTERFACE', and `FUNCTION', across different project sizes respectively (Table \ref{rq_2_1_mwstatistic}). 
\input{Tables/RQ2/succeeding_header}

Focusing on the \textbf{preceding code context of Non-Header comments}, Exception category (`EXCPTN') comments, Defintion (`DEFN') category comments, and Loops (`LOOPS') category comments have the highest SATD affliction with 21.57\%, 28.15\% and 57.36\% for project sizes 1-100 KLOC, 100-1000 KLOC and $>$1000 KLOC, respectively (Table \ref{tab:nh_prec_context}).  
Moreover, the Documentation (`DOC') category comments with 1.88\%, 1.57\%, and 0.19\% have the lowest SATD affliction across project sizes 1-100 KLOC, 100-1000 KLOC, and $>$1000 KLOC, respectively. 

In each case, hypothesis testing led to a p-value $< 0.001$, thus allowing us to reject the null hypothesis ($H_{02})$. Thus hinting that Non-Header categories `EXCPTN', `DEFN', and `LOOPS' have significant differences in SATD against their counterparts `DEFN', `EXCPTN', and `CNDTNL' for project sizes 1-100 KLOC, 100-1000 KLOC, and $>$1000 KLOC, respectively. Hence, we observe that the most common preceding source code statement category varies across project sizes, unlike header comments.

Referring to the effect size, we note that for projects between 1-100 KLOC, there is a moderate difference indicating SATD comment is more prevalent with exception handling statements as preceding context (Table \ref{rq_2_2_1_mwstatistic}). For medium-sized projects 100-1000 KLOC, the effect size is small, although statistically significant, indicating definitions as preceding context of SATD comment are relatively higher than conditional. While for large projects, the large effect size indicates loop statements have significantly more instances of SATD comments as preceding code context relative to other statement types.

\input{Tables/RQ2/preceeding_nonheader}

Finally, regarding \textbf{the succeeding code context of the Non-Header comments}, Definition category (`DEFN') comments consistently have the highest SATD affliction with 36.44\%, 27.85\%, and 42.81\% for project sizes 1-100 KLOC, 100-1000 KLOC and $>$1000 KLOC, respectively (Table~\ref{tab:nh_succ_context}).  The Loops (`LOOPS') category comments with 3.52\%, 2.60\%, and 3.09\% have the lowest SATD affliction across project sizes 1-100 KLOC, 100-1000 KLOC, and $>$1000 KLOC, respectively. 

Furthermore, in each case, hypothesis testing led to a p-value $< 0.001$, thus allowing us to reject the null hypothesis ($H_{02})$. Thus, suggesting that the Non-Header category `DEFN' has significant differences in SATD against their counterpart `CNDTNL' for project sizes 1-100 KLOC, 100-1000 KLOC, and $>$1000 KLOC, respectively. The observation is that `Definition `DEFN' is the most common succeeding source code statement category varies across project sizes, unlike header comments.

The effect size for the succeeding code context statistical evaluation of Non-Header comments suggests that for projects between 1-100 KLOC and 100-1000 KLOC, there is a moderate difference in SATD prevalence between definitions and conditional statements in smaller projects, indicating that definitions are relatively more frequent as succeeding code context of SATD comments (Table \ref{rq_2_2_2_mwstatistic}). While the same difference is more pronounced in large projects with size $>$1000 KLOC.

\input{Tables/RQ2/succeeding_nonheader}
\begin{keyRQAnswer}[\textbf{SATD Contextualization.}]
SATD co-occurs significantly more with specific code statement types.
For Header comments, CLASS statements show the highest SATD concentration (23–30\%) across all project sizes (p $< 0.001$, large effect).
For Non-Header comments, the most frequent SATD contexts are EXCPTN (small projects), DEFN (medium), and LOOPS (large) as preceding statements, and consistently DEFN as succeeding statements (p $< 0.001$).
\end{keyRQAnswer}
\subsection{$RQ_3$. SATD Frequent Code Patterns}
Tables \ref{tab:nh_freq_pattern_1_100}, \ref{tab:nh_freq_pattern_100_1000}, \ref{tab:nh_freq_pattern_gt_1000} list the cross-tabulated results of the normalized frequency percentage of preceding and succeeding code statement categories for project sizes 1-100 KLOC, 100-1000 KLOC, and $>$1000 KLOC, respectively. For projects in 1-100 KLOC, the `DEFN' as the succeeding category has the highest across almost all preceding categories, with the overall highest being 10.02\% when preceded by `BRNCH' category code statements. Evidently, it is the most frequent combination. It is followed by `EXPR' with 8.72\% as both preceding and succeeding categories, and `DECL' with 8.10\% as both preceding and succeeding categories are the second and third most frequent combination. `DOC' in both preceding and succeeding categories has the lowest percentages, indicating close to zero occurrences of SATD comments that are embedded within other comments.
\begin{table}[tb]
\caption{Non-Header SATD Patterns Frequency (\%) --- ($<100$ KLOC)}
\centering
\renewcommand{\arraystretch}{1.5}
\resizebox{\linewidth}{!}{%

\begin{tabular}{l|cccccccc}
\hline
\textbf{Succeeding} & \textbf{DECL} & \textbf{EXPR} & \textbf{DEFN} & \textbf{BRNCH} & \textbf{CNDTNL} & \textbf{LOOPS} & \textbf{EXCPTN} & \textbf{DOC} \\ 
\hline
\textbf{DECL}   & 8.10 & 2.56 & 0.08 & 0.57 & 2.63 & 0.72 & 0.21 & 0.51 \\
\textbf{EXPR}   & 2.70 & 8.72 & 2.26 & 1.48 & 2.27 & 0.56 & 0.31 & 0.68 \\
\textbf{DEFN}   & 7.04 & 4.09 & 5.03 & 10.02 & 2.32 & 0.35 & 1.00 & 2.37 \\
\textbf{BRNCH}  & 0.12 & 0.03 & 1.00 & 0.05 & 0.41 & 0.01 & 0.02 & 0.30 \\
\textbf{CNDTNL} & 4.58 & 6.53 & 0.69 & 3.03 & 4.22 & 0.34 & 0.86 & 0.24 \\
\textbf{LOOPS}  & 0.73 & 0.81 & 0.06 & 0.18 & 0.58 & 0.22 & 0.05 & 0.03 \\
\textbf{EXCPTN} & 0.66 & 6.36 & 0.11 & 0.37 & 0.32 & 0.05 & 0.42 & 0.07 \\
\textbf{DOC}    & 0.00 & 0.03 & 0.01 & 0.00 & 0.00 & 0.00 & 0.00 & 0.00 \\
\hline
\end{tabular}%

}
\label{tab:nh_freq_pattern_1_100}
\end{table}

\begin{table}[tb]
\caption{Non-Header SATD Patterns Frequency (\%) --- ($100-1000$ KLOC)}
\centering
\renewcommand{\arraystretch}{1.5}
\resizebox{\linewidth}{!}{%
\begin{tabular}{l|cccccccc}
\hline
\textbf{Succeeding} & \textbf{DECL} & \textbf{EXPR} & \textbf{DEFN} & \textbf{BRNCH} & \textbf{CNDTNL} & \textbf{LOOPS} & \textbf{EXCPTN} & \textbf{DOC} \\ 
\hline
\textbf{DECL}   & 7.30 & 2.35 & 0.08 & 0.55 & 2.50 & 0.63 & 0.16 & 1.28 \\
\textbf{EXPR}   & 2.72 & 7.59 & 1.37 & 1.09 & 2.29 & 0.39 & 0.48 & 1.02 \\
\textbf{DEFN}   & 6.25 & 5.60 & 6.73 & 10.89 & 2.33 & 0.27 & 1.44 & 4.88 \\
\textbf{BRNCH}  & 0.12 & 0.04 & 0.59 & 0.06 & 0.57 & 0.00 & 0.03 & 0.25 \\
\textbf{CNDTNL} & 4.10 & 6.50 & 0.19 & 2.84 & 3.92 & 0.43 & 1.05 & 0.27 \\
\textbf{LOOPS}  & 0.67 & 0.69 & 0.05 & 0.16 & 0.55 & 0.16 & 0.05 & 0.04 \\
\textbf{EXCPTN} & 0.57 & 4.49 & 0.28 & 0.27 & 0.37 & 0.02 & 0.42 & 0.10 \\
\textbf{DOC}    & 0.00 & 0.00 & 0.01 & 0.00 & 0.01 & 0.00 & 0.00 & 0.00 \\
\hline
\end{tabular}%
}
\label{tab:nh_freq_pattern_100_1000}
\end{table}

For projects sized between 100-1000 KLOC, `DEFN' as the succeeding category remains prominent across most preceding categories, with the highest percentage observed when preceded by `BRNCH' at 10.89\%. The second most common pattern involves `EXPR' as both preceding and succeeding categories. The `DECL' category as bothe preceding and succeeding categories is the third most common pattern with 7.30\%. The `DOC' category as succeeding category remains the lowest across all preceding categories followed by the `LOOPS' and `BRNCH' categories.

\begin{table}[tb]
\scalefont{0.82} 
\caption{Non-Header SATD Patterns Frequency (\%) --- ($>1000$ KLOC)}
\centering
\renewcommand{\arraystretch}{1.5}
\resizebox{\linewidth}{!}{%
\begin{tabular}{l|cccccccc}
\hline
\textbf{Succeeding} & \textbf{DECL} & \textbf{EXPR} & \textbf{DEFN} & \textbf{BRNCH} & \textbf{CNDTNL} & \textbf{LOOPS} & \textbf{EXCPTN} & \textbf{DOC} \\ 
\hline
\textbf{DECL}   & 7.93 & 2.65 & 0.18 & 0.63 & 3.06 & 0.35 & 0.17 & 0.02 \\
\textbf{EXPR}   & 2.34 & 7.34 & 1.03 & 1.01 & 2.42 & 0.36 & 0.18 & 1.01 \\
\textbf{DEFN}   & 6.30 & 3.88 & 4.25 & 8.74 & 3.39 & 0.38 & 2.12 & 2.61 \\
\textbf{BRNCH}  & 0.05 & 0.27 & 0.81 & 0.14 & 0.89 & 0.02 & 0.00 & 1.47 \\
\textbf{CNDTNL} & 5.49 & 7.78 & 0.15 & 3.83 & 5.73 & 0.35 & 1.54 & 0.28 \\
\textbf{LOOPS}  & 0.53 & 0.63 & 0.02 & 0.60 & 0.65 & 0.11 & 0.05 & 0.12 \\
\textbf{EXCPTN} & 0.53 & 3.13 & 0.08 & 0.82 & 0.48 & 0.02 & 0.94 & 0.16 \\
\textbf{DOC}    & 0.01 & 0.00 & 0.59 & 0.00 & 0.00 & 0.00 & 0.00 & 0.00 \\
\hline
\end{tabular}%
}
\label{tab:nh_freq_pattern_gt_1000}
\end{table}

For projects over 1000 KLOC, the `DEFN' category consistently is the most frequent category, with `BRNCH' as the preceding category. This combination is the most frequent pattern with 8.74\% in larger projects as well. The `DECL' as both preceding and succeeding categories is the second most common SATD pattern with 7.93\%. Unlike other project sizes, `CNDTNL' as succeeding category and `EXPR' as preceding category is the third most frequent pattern with 7.78\%. Similar to the other size categories, `DOC' remains the least common context for SATD comments, suggesting that, regardless of project size, technical debt is less often admitted within documentation comments.

%% file: Tables/RQ1/characterization.tex
\begin{table}[tb]
\footnotesize
\caption{SATD Localization Distribution}
  \label{tab:localization_statistical_test}
\resizebox{\linewidth}{!}{
\begin{tabular}{c|c|ccccc}
\hline
\multirow{3}{*}{\textbf{\begin{tabular}[c]{@{}c@{}}Project \\ Size \\(KLOC)\end{tabular}}} &
  \multirow{3}{*}{\textbf{\begin{tabular}[c]{@{}c@{}}\# \\ Projects \end{tabular}}} &
  \multicolumn{5}{c}{\textbf{\% Normalized Proportion}} \\ \cline{3-7} 
 &
   &
  \multicolumn{2}{c|}{\textbf{\begin{tabular}[c]{@{}c@{}}Header \\ SATD\end{tabular}}} &
  \multicolumn{2}{c|}{\textbf{\begin{tabular}[c]{@{}c@{}}Non-Header \\ SATD\end{tabular}}} &
  \multirow{2}{*}{\textbf{p-value}} \\ \cline{3-6}
 &
   &
  \multicolumn{1}{c|}{\textbf{Mean}} &
  \multicolumn{1}{c|}{\textbf{Median}} &
  \multicolumn{1}{c|}{\textbf{Mean}} &
  \multicolumn{1}{c|}{\textbf{Median}} &
   \\ \hline
\begin{tabular}[c]{@{}c@{}}1-100\end{tabular} &
  4,752 &
  \multicolumn{1}{c|}{0.025} &
  \multicolumn{1}{c|}{0.003} &
  \multicolumn{1}{c|}{0.070} &
  \multicolumn{1}{c|}{0.031} &
  \textless 0.001 \\
\begin{tabular}[c]{@{}c@{}}100-1000\end{tabular} &
  527 &
  \multicolumn{1}{c|}{0.009} &
  \multicolumn{1}{c|}{0.003} &
  \multicolumn{1}{c|}{0.038} &
  \multicolumn{1}{c|}{0.022} &
  \textless 0.001 \\
\begin{tabular}[c]{@{}c@{}}\textgreater{}1000 \end{tabular} &
  41 &
  \multicolumn{1}{c|}{0.005} &
  \multicolumn{1}{c|}{0.003} &
  \multicolumn{1}{c|}{0.019} &
  \multicolumn{1}{c|}{0.010} &
  \textless 0.001 \\ \hline
\end{tabular}
}
\end{table}

%% file: Tables/RQ1/h01.tex
\begin{table}[tb]
\centering
\caption{SATD Localization Distribution ($H_{01}$)}
\resizebox{\linewidth}{!}{
\begin{tabular}{m{1.5cm}| l l l l}
\hline
\textbf{Project size [KLOC]} & \textbf{\# Projects} & \textbf{U-Statistic} & \textbf{R Coefficient} & \textbf{Effect Size} \\
\hline
1-100 & 4752 & 4852343.0 & 0.570 & Large \\
100-1000 & 527 & 54189.5 & 0.609 & Large \\ 
$>$1000 & 41 & 337.0 & 0.599 & Large \\
\hline
\end{tabular}
}
\label{rq_1_mwstatistic}
\end{table}

%% file: Tables/RQ2/succeeding_header.tex
\begin{table}[tb]
\caption{Header-$N_S$ Distribution}
\centering
\begin{tabular}{m{2cm} |m{1.2cm} m{1.2cm} m{1.2cm}}
\hline
\multirow{2}{*}{\textbf{Category}}
 & \multicolumn{3}{c}{\textbf{KLOC}}\\ \cline{2-4}
 & \textbf{1-100} & \textbf{100-1000} & \textbf{1000} \\ \hline
CLASS & 23.22 & 26.54 & 30.15 \\

CONSTRUCTOR & 22.01 & 9.25 & 7.71 \\

ENUM & 19.67 & 18.0 & 15.0 \\

FILE & 0.72 & 4.53 & 4.52 \\

FUNCTION & 19.89 & 20.05 & 25.88 \\

INTERFACE & 14.49 & 21.64 & 16.73 \\
\hline
\end{tabular}
\label{tab:h_succ_context}
\end{table}

\begin{table}[tb]
\footnotesize
\caption{Header-$N_S$ Distribution by SATD Type}
\centering
 \begin{tabular}{m{2.2cm} |m{1cm} m{1.5cm} m{2cm}}
\hline
\textbf{Statement Category}
 & \textbf{ETD }& \textbf{HTD (noisy)} & \textbf{HTD (validated)} \\ \hline
CLASS & 23.22 & 26.54 & 30.15 \\

CONSTRUCTOR & 22.01 & 9.25 & 7.71 \\

ENUM & 19.67 & 18.0 & 15.0 \\

FILE & 0.72 & 4.53 & 4.52 \\

FUNCTION & 19.89 & 20.05 & 25.88 \\

INTERFACE & 14.49 & 21.64 & 16.73 \\
\hline
\end{tabular}
\label{tab:h_succ_context_2}
\end{table}

\begin{table}[tb]
\centering
\caption{Header-$N_S$ Distribution ($H_{02}$)}
\label{rq_2_1_mwstatistic}
\resizebox{\linewidth}{!}{
\begin{tabular}{m{1.3cm} |l l l l}
\hline
\textbf{Project [KLOC]} & \textbf{\# Projects} & \textbf{U-Statistic} & R \textbf{Coefficient} & \textbf{Effect Size} \\
\hline
1-100 & 2710 & 5262982.0 & 0.433 & Large \\
100-1000 & 482 & 191255.5 & 0.646 & Large \\ 
$>$1000 & 38 & 425.0 & 0.411 & Large \\
\hline
\end{tabular}

}
\end{table}

%% file: Tables/RQ2/preceeding_nonheader.tex
\begin{table}[tb]
\caption{Non-Header-$N_P$ Distribution}
\centering
\begin{tabular}{l |m{1cm}  m{1.1cm} m{1cm}}
\hline
\multirow{2}{*}{\textbf{Category}}
 & \multicolumn{3}{c}{\textbf{KLOC}}\\ \cline{2-4}
 & \textbf{1-100} & \textbf{100-1000} & \textbf{1000} \\ \hline
DECLARATION (DECL) & 10.21 & 10.44 & 0.93 \\

EXPRESSION (EXPR) & 5.75 & 3.81 & 1.76 \\

DEFINITION (DEFN) & 18.90 & 28.15 & 4.67 \\

BRANCHING (BRNCH) & 10.98 & 11.11 & 4.74 \\

CONDITIONAL (CNDTNL) & 19.49 & 21.00 & 25.05 \\

LOOPS (LOOPS) & 11.22 & 3.65 & 57.36\\

EXCEPTION (EXCPTN) & 21.57 & 20.28 & 5.30 \\

DOCUMENTATION (DOC) & 1.88 & 1.57 & 0.19 \\
\hline
\end{tabular}
\label{tab:nh_prec_context}
\end{table}

\begin{table}[tb]
\centering
\caption{Non-Header-$N_P$ Distribution ($H_{02}$)}
\label{rq_2_2_1_mwstatistic}
\resizebox{\linewidth}{!}{
\begin{tabular}{m{1.3cm}| l l l l}
\hline
\textbf{Project [KLOC]} & \textbf{\# Projects} & \textbf{U-Statistic} & R \textbf{Coefficient} & \textbf{Effect Size }\\
\hline
1-100 & 4496 & 7043822.5 & 0.303 & Medium \\
100-1000 & 512 & 156303.0 & 0.19 & Small \\ 
$>$1000 & 41 & 315.0 & 0.625 & Large \\
\hline
\end{tabular}

}
\end{table}

%% file: Tables/RQ2/succeeding_nonheader.tex
\begin{table}[tb]
\caption{Non-Header-$N_S$ Distribution}
\centering

\begin{tabular}{l |m{1cm} m{1.1cm} m{1cm}}
\hline
\multirow{2}{*}{\textbf{Category}}
 & \multicolumn{3}{c}{\textbf{KLOC}}\\ \cline{2-4}
 & \textbf{1-100} & \textbf{100-1000} & \textbf{1000} \\ \hline
DECLARATION (DECL) & 7.05 & 6.95 & 1.52 \\

EXPRESSION (EXPR) & 5.23 & 3.65 & 3.19 \\

DEFINITION (DEFN) & 36.44 & 27.85 & 42.81 \\

BRANCHING (BRNCH) & 13.72 & 11.16 & 10.73 \\

CONDITIONAL (CNDTNL) & 15.96 & 25.64 & 16.67 \\

LOOPS (LOOPS) & 3.52 & 2.60& 3.09 \\

EXCEPTION (EXCPTN) & 10.98 & 10.80 & 14.21 \\

DOCUMENTATION (DOC) & 7.09 & 11.35 & 7.80 \\
\hline
\end{tabular}
\label{tab:nh_succ_context}
\end{table}

\begin{table}[tb]
\centering
\caption{Non-Header-$N_S$ Distribution ($H_{02}$)}
\begin{tabular}{m{1.3cm}| l l l l}
\hline
\textbf{Project [KLOC] }& \textbf{\# Projects} & \textbf{U-Statistic} & \textbf{R Coefficient }& \textbf{Effect Size }\\
\hline
1-100 & 4496 & 8082477.5 & 0.200 & Medium \\
100-1000 & 512 & 90903.5 & 0.306 & Medium \\ 
$>$1000 & 41 & 406.6.0 & 0.516 & Large \\
\hline
\end{tabular}
\label{rq_2_2_2_mwstatistic}
\end{table}

%% file: sections/discussions.tex
The large and consistent effect sizes (ranging from 0.57 to 0.61 across project sizes) confirm that the differences in SATD prevalence between Non-Header and Header comments are not only statistically significant but also practically meaningful. This finding highlights a structural tendency in how developers communicate debt: SATD is overwhelmingly anchored in the executable regions of code, rather than in declarative or documentation-level segments. Consequently, the locus of SATD aligns with where developers make implementation decisions and trade-offs, suggesting that SATD often originates from situated, fine-grained reasoning within development activities rather than from high-level design rationale. Understanding this behavioral pattern has direct implications for how both researchers and practitioners conceptualize, detect, and manage TD.

\begin{keyTakeAways}[SATD Localization ($RQ_1$)]
Our findings reposition SATD as an \emph{implementation-anchored phenomenon}. For practitioners, this means SATD detection and monitoring tools should prioritize scanning and interpreting inline and block comments within method bodies, where most self-admissions occur. For researchers, this calls for a shift in focus from broad file- or class-level analyses to finer-grained localization models that capture the dynamic interplay between code evolution and self-admitted concerns.
\end{keyTakeAways}

From an architectural perspective, the consistent concentration of SATD in `CLASS' Header comments suggests that debt at the class level reflects structural and design-level complexity rather than simple documentation issues. The rarity of `FILE' Header SATD underscores that TD is perceived and documented closer to functional boundaries than to organizational units such as files. The observed decline of `CONSTRUCTOR'-related SATD in larger projects, contrasted with a rise in `FUNCTION'-related debt, suggests that as systems grow, developers increasingly confront debt in more complex behavioral logic rather than in initialization code. This evolution indicates a maturation of design practices, but also a shift in the debt landscape toward maintainability and scalability challenges.

\begin{keyTakeAways}[SATD Code Contextualization ($RQ_2$)]
The contextual analysis reveals that SATD is not uniformly distributed across the codebase but is structurally coupled with specific constructs that embody uncertainty, decision-making, or error management. Practitioners can leverage these insights to design \emph{context-aware debt monitoring} that targets high-risk regions such as conditionals, branching, and exception handling. For researchers, these contextual signatures provide a foundation for predictive models of SATD formation and propagation, integrating structural features with historical and textual cues.
\end{keyTakeAways}

The recurring patterns identified across project sizes show that definitions (`DEFN`) and branching (`BRNCH`) frequently co-occur with SATD. This finding suggests that developers are most likely to express debt when defining new entities or making conditional decisions that affect control flow, points where trade-offs and compromises are most visible. The dominance of `DEFN` as a succeeding context across all project sizes indicates that TD tends to crystallize immediately before the introduction of new functionality or abstractions. This pattern offers a novel insight: SATD is often a precursor to structural expansion rather than a byproduct of it. In other words, developers “admit debt” just before extending or refactoring code, turning SATD into a cognitive checkpoint where awareness and intent intersect.

\begin{keyTakeAways}[SATD Frequent Code Patterns ($RQ_3$)]
For practitioners, the recurring patterns suggest where automated detection or refactoring support could be most effective: in code areas involving definitions, branching, and exception handling. These insights can inform IDE-integrated “SATD hotspots” that warn developers when they are operating in contexts historically prone to debt admission. For researchers, the identified co-occurrence patterns open avenues for modeling SATD emergence as a structural phenomenon, enabling future tools to reason not just about \textbf{what} is admitted as debt, \textbf{but when and where} it tends to occur in the code lifecycle.
\end{keyTakeAways}


%% file: sections/validity.tex
In this section, we discuss the threats to Validity of our study, following the guidelines defined by Wohlin et al.~\cite{wohlin_experimentation_2024}.

\textbf{Internal Validity.}
A potential threat to internal validity lies in the quality and accuracy of the dataset used. 
Our analysis relies on the PENTACET dataset~\cite{sridharan2023pentacet}, which automatically identified over 250{,}000 SATD instances across more than 9{,}000 Java repositories. 
The dataset differentiates between Easy-To-Find (ETF) and Hard-To-Find (HTF) SATD instances. While the original study reported an Inter-Rater Agreement (\textbf{IRA}) of $\kappa=0.75$ for HTF instances, indicating substantial reliability, it did not provide a corresponding value for ETF instances. To address this gap, we also performed an independent validation on a randomly selected subsample of ETF and non-SATD comments, determined using a 95\% confidence level and a 5\% margin of error. The validation was carried out by two authors independently who manually reviewed the selected comments. The resulting Cohen’s $\kappa = 0.96$ demonstrates almost perfect agreement, thereby reinforcing the reliability and consistency of the dataset’s annotations.

\textbf{Construct Validity.}
Construct validity threats may arise if the operationalized constructs fail to reflect their intended theoretical meaning. 
In our study, the categorization of statement types surrounding SATD comments could have introduced ambiguity if not grounded in systematic semantics. 
To minimize such risks, statement types were organized into nine coherent categories (Section~\ref{sec:dc}) based on the \textit{srcML}~\cite{collard2013srcml} documentation, the same parser used in compiling the PENTACET dataset, and corroborated through the official Java documentation\footnote{\url{https://docs.oracle.com/javase}}. 
This dual-source grounding ensures that our classification reflects the language’s syntactic and semantic structures rather than researcher's interpretation. 
Explicit inclusion and exclusion criteria were defined for all statement types, ensuring replicability and minimizing coder bias in operationalization. 

\textbf{External Validity.}
Our empirical analysis is based on Java projects of diverse sizes and domains, providing a broad and representative sample within that ecosystem. 
Nevertheless, results may not directly generalize to other programming languages or paradigms, where comment practices, syntactic constructs, and technical debt disclosure norms differ. 
To address this limitation, our methodology was intentionally designed to be \textit{language-agnostic}, relying on abstract syntax tree representations derived from \textit{srcML}. 
This design allows our approach to be replicated in other ecosystems by substituting language-specific grammars. 
While cross-language replication would be required to confirm generalizability, the study provides a reusable framework for analyzing the structural and contextual manifestations of SATD in source code.

\textbf{Conclusion Validity.}
Threats to conclusion validity concern the reliability of statistical inferences and the adequacy of the analytical procedures used. 
We applied the non-parametric Wilcoxon–Mann–Whitney U test to evaluate differences in SATD prevalence and contextual distributions, following recommended practices for non-normal data. 
Effect sizes were computed using the rank-biserial correlation coefficient ($R$) and interpreted according to Kampenes et al.~\cite{kampenes2007systematic}, allowing us to assess not only statistical but also the magnitude of the significance. 
All tests were performed with a significance level ($\alpha$) of 0.001. 
The large number of independent projects mitigates risks of sampling bias and supports the robustness of our conclusions. 
However, as with any large-scale dataset, potential dependencies among data points (e.g., projects sharing codebases or contributors) could still influence the true independence of observations. 
We mitigated this by analyzing projects individually and aggregating results at the project level rather than per comment, thereby reducing the impact of intra-project correlation.

%% file: sections/conclusion.tex
Our study tries to bridge the gap between SATD and its surrounding source code, revealing where and how developers express debt in practice. By analyzing over 225,000 SATD instances from the PENTACET dataset, we found that SATD is mainly rooted in non-header (inline) code, where developers make everyday implementation decisions rather than in documentation or design sections.

SATD most often appears near definitions, conditionals, branching, and exception handling, places where uncertainty and trade-offs are high. These patterns show that developers tend to admit debt right before extending or modifying functionality, turning SATD into a natural checkpoint in the coding process.
Our findings suggest that debt detection tools should move beyond keywords to focus on context-aware analysis, targeting the code areas most likely to accumulate debt. Future work should explore how these patterns hold across programming languages, evolve, and inform AI-driven tools that predict or mitigate debt before it grows.

%% file: main.bbl
\begin{thebibliography}{10}
\providecommand{\url}[1]{#1}
\csname url@samestyle\endcsname
\providecommand{\newblock}{\relax}
\providecommand{\bibinfo}[2]{#2}
\providecommand{\BIBentrySTDinterwordspacing}{\spaceskip=0pt\relax}
\providecommand{\BIBentryALTinterwordstretchfactor}{4}
\providecommand{\BIBentryALTinterwordspacing}{\spaceskip=\fontdimen2\font plus
\BIBentryALTinterwordstretchfactor\fontdimen3\font minus \fontdimen4\font\relax}
\providecommand{\BIBforeignlanguage}[2]{{%
\expandafter\ifx\csname l@#1\endcsname\relax
\typeout{** WARNING: IEEEtran.bst: No hyphenation pattern has been}%
\typeout{** loaded for the language `#1'. Using the pattern for}%
\typeout{** the default language instead.}%
\else
\language=\csname l@#1\endcsname
\fi
#2}}
\providecommand{\BIBdecl}{\relax}
\BIBdecl

\bibitem{cunningham1992wycash}
W.~Cunningham, ``The wycash portfolio management system,'' \emph{ACM SIGPLAN OOPS Messenger}, vol.~4, no.~2, pp. 29--30, 1992.

\bibitem{li2015systematic}
Z.~Li, P.~Avgeriou, and P.~Liang, ``A systematic mapping study on technical debt and its management,'' \emph{Journal of Systems and Software}, vol. 101, pp. 193--220, 2015.

\bibitem{potdar2014exploratory}
A.~Potdar and E.~Shihab, ``An exploratory study on self-admitted technical debt,'' in \emph{2014 IEEE International Conference on Software Maintenance and Evolution}.\hskip 1em plus 0.5em minus 0.4em\relax IEEE, 2014, pp. 91--100.

\bibitem{huang2018identifying}
Q.~Huang, E.~Shihab, X.~Xia, D.~Lo, and S.~Li, ``Identifying self-admitted technical debt in open source projects using text mining,'' \emph{Empirical Software Engineering}, vol.~23, no.~1, pp. 418--451, 2018.

\bibitem{de2020identifying}
M.~A. de~Freitas~Farias, M.~G. de~Mendon{\c{c}}a~Neto, M.~Kalinowski, and R.~O. Sp{\'\i}nola, ``Identifying self-admitted technical debt through code comment analysis with a contextualized vocabulary,'' \emph{Information and Software Technology}, vol. 121, p. 106270, 2020.

\bibitem{sridharan2021data}
M.~Sridharan, M.~Mantyla, L.~Rantala, and M.~Claes, ``Data balancing improves self-admitted technical debt detection,'' in \emph{2021 IEEE/ACM 18th International Conference on Mining Software Repositories (MSR)}.\hskip 1em plus 0.5em minus 0.4em\relax IEEE, 2021, pp. 358--368.

\bibitem{alhefdhi2022framework}
A.~Alhefdhi, H.~K. Dam, Y.~S. Nugroho, H.~Hata, T.~Ishio, and A.~Ghose, ``A framework for conditional statement technical debt identification and description,'' \emph{Automated Software Engineering}, vol.~29, no.~2, p.~60, 2022.

\bibitem{alomar2022satdbailiff}
E.~A. AlOmar, B.~Christians, M.~Busho, A.~H. AlKhalid, A.~Ouni, C.~Newman, and M.~W. Mkaouer, ``Satdbailiff-mining and tracking self-admitted technical debt,'' \emph{Science of Computer Programming}, vol. 213, p. 102693, 2022.

\bibitem{maldonado2017empirical}
E.~D.~S. Maldonado, R.~Abdalkareem, E.~Shihab, and A.~Serebrenik, ``An empirical study on the removal of self-admitted technical debt,'' in \emph{2017 IEEE International Conference on Software Maintenance and Evolution (ICSME)}.\hskip 1em plus 0.5em minus 0.4em\relax IEEE, 2017, pp. 238--248.

\bibitem{liu2018satd}
Z.~Liu, Q.~Huang, X.~Xia, E.~Shihab, D.~Lo, and S.~Li, ``Satd detector: A text-mining-based self-admitted technical debt detection tool,'' in \emph{Proceedings of the 40th International Conference on Software Engineering: Companion Proceeedings}, 2018, pp. 9--12.

\bibitem{de2005study}
S.~C.~B. de~Souza, N.~Anquetil, and K.~M. de~Oliveira, ``A study of the documentation essential to software maintenance,'' in \emph{Proceedings of the 23rd Annual International Conference on Design of Communication: documenting \& designing for pervasive information}, 2005, pp. 68--75.

\bibitem{woodfield1981effect}
S.~N. Woodfield, H.~E. Dunsmore, and V.~Y. Shen, ``The effect of modularization and comments on program comprehension,'' in \emph{Proceedings of the 5th international conference on Software engineering}, 1981, pp. 215--223.

\bibitem{tenny1988program}
T.~Tenny, ``Program readability: Procedures versus comments,'' \emph{IEEE Transactions on Software Engineering}, vol.~14, no.~9, pp. 1271--1279, 1988.

\bibitem{miyake2017replicated}
Y.~Miyake, S.~Amasaki, H.~Aman, and T.~Yokogawa, ``A replicated study on relationship between code quality and method comments,'' \emph{Applied computing and information technology}, pp. 17--30, 2017.

\bibitem{padioleau2009listening}
Y.~Padioleau, L.~Tan, and Y.~Zhou, ``Listening to programmers—taxonomies and characteristics of comments in operating system code,'' in \emph{2009 IEEE 31st International Conference on Software Engineering}.\hskip 1em plus 0.5em minus 0.4em\relax IEEE, 2009, pp. 331--341.

\bibitem{haouari2011good}
D.~Haouari, H.~Sahraoui, and P.~Langlais, ``How good is your comment? a study of comments in java programs,'' in \emph{2011 International Symposium on Empirical Software Engineering and Measurement}.\hskip 1em plus 0.5em minus 0.4em\relax IEEE, 2011, pp. 137--146.

\bibitem{steidl2013quality}
D.~Steidl, B.~Hummel, and E.~Juergens, ``Quality analysis of source code comments,'' in \emph{2013 21st international conference on program comprehension (icpc)}.\hskip 1em plus 0.5em minus 0.4em\relax Ieee, 2013, pp. 83--92.

\bibitem{pascarella2017classifying}
L.~Pascarella and A.~Bacchelli, ``Classifying code comments in java open-source software systems,'' in \emph{2017 IEEE/ACM 14th International Conference on Mining Software Repositories (MSR)}.\hskip 1em plus 0.5em minus 0.4em\relax IEEE, 2017, pp. 227--237.

\bibitem{sridhara2010towards}
G.~Sridhara, E.~Hill, D.~Muppaneni, L.~Pollock, and K.~Vijay-Shanker, ``Towards automatically generating summary comments for java methods,'' in \emph{Proceedings of the 25th IEEE/ACM international conference on Automated software engineering}, 2010, pp. 43--52.

\bibitem{vassallo2014codes}
C.~Vassallo, S.~Panichella, M.~Di~Penta, and G.~Canfora, ``Codes: Mining source code descriptions from developers discussions,'' in \emph{Proceedings of the 22nd International Conference on Program Comprehension}, 2014, pp. 106--109.

\bibitem{malik2008understanding}
H.~Malik, I.~Chowdhury, H.-M. Tsou, Z.~M. Jiang, and A.~E. Hassan, ``Understanding the rationale for updating a function’s comment,'' in \emph{2008 IEEE International Conference on Software Maintenance}.\hskip 1em plus 0.5em minus 0.4em\relax IEEE, 2008, pp. 167--176.

\bibitem{chen2019automatically}
H.~Chen, Y.~Huang, Z.~Liu, X.~Chen, F.~Zhou, and X.~Luo, ``Automatically detecting the scopes of source code comments,'' \emph{Journal of Systems and Software}, vol. 153, pp. 45--63, 2019.

\bibitem{fenton1998software}
N.~E. Fenton and S.~L. Pfleeger, ``Software metrics: A rigorous and practical approach: Brooks,'' 1998.

\bibitem{basili2002validation}
V.~R. Basili, L.~C. Briand, and W.~L. Melo, ``A validation of object-oriented design metrics as quality indicators,'' \emph{IEEE Transactions on software engineering}, vol.~22, no.~10, pp. 751--761, 2002.

\bibitem{AlOmar2022}
E.~A. AlOmar, B.~Christians, M.~Busho, A.~H. AlKhalid, A.~Ouni, C.~D. Newman, and M.~W. Mkaouer, ``{SATDBailiff}: Mining and tracking self-admitted technical debt,'' \emph{Science of Computer Programming}, vol. 213, p. 102693, 2022.

\bibitem{Sridharan2023}
M.~Sridharan, L.~Rantala, and M.~M{"a}ntyl{"a}, ``{PENTACET} data – 23 million contextual code comments and 250,000 {SATD} comments,'' in \emph{Proceedings of the 20th ACM/IEEE International Conference on Mining Software Repositories (MSR) - Data/Tool Track}.\hskip 1em plus 0.5em minus 0.4em\relax IEEE, 2023, doi: 10.5281/zenodo.7767734.

\bibitem{Pham2025}
P.~Pham, M.~Sridharan, M.~Esposito, and V.~Lenarduzzi, ``Descriptor: {C++} self-admitted technical debt dataset (cppsatd),'' \emph{IEEE Data Descriptions}, 2025, (to appear).

\bibitem{Zampetti2018}
F.~Zampetti, A.~Serebrenik, and M.~Di~Penta, ``Was self-admitted technical debt removal a real removal? an in-depth perspective,'' in \emph{Proceedings of the 15th International Conference on Mining Software Repositories (MSR)}.\hskip 1em plus 0.5em minus 0.4em\relax ACM, 2018, p. 526–536.

\bibitem{Alhefdhi2024}
A.~Alhefdhi, H.~K. Dam, and A.~Ghose, ``Towards automating self-admitted technical debt repayment,'' \emph{Information and Software Technology}, vol. 167, p. 107376, 2024.

\bibitem{Yonekura2025}
M.~Yonekura, Y.~Kashiwa, B.~Lin, K.~Fujiwara, and H.~Iida, ``Leveraging context information for self-admitted technical debt detection,'' in \emph{Proceedings of the 33rd IEEE/ACM International Conference on Program Comprehension (ICPC)}, 2025.

\bibitem{wohlin_experimentation_2024}
\BIBentryALTinterwordspacing
C.~Wohlin, P.~Runeson, M.~Höst, M.~C. Ohlsson, B.~Regnell, and A.~Wesslén, \emph{Experimentation in {Software} {Engineering}, {Second} {Edition}}.\hskip 1em plus 0.5em minus 0.4em\relax Springer, 2024. [Online]. Available: \url{https://doi.org/10.1007/978-3-662-69306-3}
\BIBentrySTDinterwordspacing

\bibitem{de2016investigating}
M.~A. de~Freitas~Farias, J.~A. Santos, M.~Kalinowski, M.~Mendon{\c{c}}a, and R.~O. Sp{\'\i}nola, ``Investigating the identification of technical debt through code comment analysis,'' in \emph{International Conference on Enterprise Information Systems}.\hskip 1em plus 0.5em minus 0.4em\relax Springer, 2016, pp. 284--309.

\bibitem{de2016identifying}
M.~A. de~Freitas~Farias, M.~Cola{\c{c}}o, R.~O. Sp{\'\i}nola, and M.~G. de~Mendon{\c{c}}a~Neto, ``Identifying technical debt through code comment analysis,'' in \emph{Doctoral Consortium on Enterprise Information Systems}, vol.~2.\hskip 1em plus 0.5em minus 0.4em\relax SCITEPRESS, 2016.

\bibitem{sridharan2023pentacet}
M.~Sridharan, L.~Rantala, and M.~M{\"a}ntyl{\"a}, ``Pentacet data-23 million contextual code comments and 250,000 satd comments,'' in \emph{2023 IEEE/ACM 20th International Conference on Mining Software Repositories (MSR)}.\hskip 1em plus 0.5em minus 0.4em\relax IEEE, 2023, pp. 412--416.

\bibitem{collard2013srcml}
M.~L. Collard, M.~J. Decker, and J.~I. Maletic, ``srcml: An infrastructure for the exploration, analysis, and manipulation of source code: A tool demonstration,'' in \emph{2013 IEEE International conference on software maintenance}.\hskip 1em plus 0.5em minus 0.4em\relax IEEE, 2013, pp. 516--519.

\bibitem{falessi_enhancing_2023}
D.~Falessi, S.~M. Laureani, J.~Çarka, M.~Esposito, and D.~A.~d. Costa, ``Enhancing the defectiveness prediction of methods and classes via {JIT},'' \emph{Empir. Softw. Eng.}, vol.~28, no.~2, p.~37, 2023.

\bibitem{anderson1952asymptotic}
T.~W. Anderson and D.~A. Darling, ``Asymptotic theory of certain" goodness of fit" criteria based on stochastic processes,'' \emph{The annals of mathematical statistics}, pp. 193--212, 1952.

\bibitem{mann1947test}
H.~B. Mann and D.~R. Whitney, ``On a test of whether one of two random variables is stochastically larger than the other,'' \emph{The annals of mathematical statistics}, pp. 50--60, 1947.

\bibitem{cureton1956rank}
E.~E. Cureton, ``Rank-biserial correlation,'' \emph{Psychometrika}, vol.~21, no.~3, pp. 287--290, 1956.

\bibitem{kampenes2007systematic}
V.~B. Kampenes, T.~Dyb{\aa}, J.~E. Hannay, and D.~I. Sj{\o}berg, ``A systematic review of effect size in software engineering experiments,'' \emph{Information and Software Technology}, vol.~49, no. 11-12, pp. 1073--1086, 2007.

\end{thebibliography}
